\documentclass[prl,superscriptaddress,twocolumn,floatfix,showpacs,letterpaper]{revtex4}
\usepackage{amsmath}
\usepackage{graphicx}
\usepackage{times}
\usepackage{hyperref}
\usepackage{subfigure}

\begin{document}
\title{Fundamental high pressure calibration from all-electron quantum
  Monte Carlo calculations} \date{\today}

\author{K. P. Esler}
\affiliation{Carnegie Institution of Washington, Geophysical
  Laboratory}
\affiliation{University of Illinois at Urbana-Champaign, NCSA}
\email{esler@uiuc.edu}
\author{R. E. Cohen}
\affiliation{Carnegie Institution of Washington, Geophysical Laboratory}
\author{B. Militzer}
\affiliation{University of California, Berkeley, Dept. of Earth and
  Planetary Science and of Astronomy}
\author{Jeongnim Kim}
\affiliation{University of Illinois at Urbana-Champaign, NCSA}
\author{R.J. Needs}
\affiliation{Theory of Condensed Matter Group, Cavendish Laboratory,
  Cambridge CB3 0HE, United Kingdom}
\author{M.D. Towler}
\affiliation{Theory of Condensed Matter Group, Cavendish Laboratory,
  Cambridge CB3 0HE, United Kingdom}
\begin{abstract}
We develop an all-electron quantum Monte Carlo (QMC) method for solids
that does not rely on pseudopotentials, and use it to construct a
primary ultra-high pressure calibration based the equation of state of
cubic boron nitride(c-BN).  We compute the static contribution to the
free energy with QMC, and obtain the phonon contribution from density
functional theory, yielding a high-accuracy calibration up to 900 GPa
usable directly in experiment.  Furthermore, we compute the anharmonic
Raman frequency shift with QMC as a function of pressure and
temperature, allowing optical pressure calibration in table-top
experiments.  In contrast to present experimental approaches, small
systematic errors in the theoretical EOS do not increase with
pressure, and no extrapolation is needed.  This all-electron
methodology is generally applicable to first-row solids, and can be
used to provide a new reference for {\em ab initio} calculations of
solids and to benchmark pseudopotential accuracy.
\end{abstract}
\maketitle

Although the number of studies of materials using density functional
theory (DFT) continues to grow explosively, the accuracy of their
predictions is variable, sometimes excellent and sometimes poor,
limiting the confidence one may place in DFT results as a quantitative
calibration for experimental studies.  Quantum Monte Carlo (QMC),
particularly diffusion Monte Carlo (DMC), is the highest-accuracy
method for finding the ground state of a many-electron Hamiltonian.
For solids with atoms heavier than He, however, the Hamiltonian itself
is approximated using pseudopotentials (PPs) based on a lower-accuracy
theory, limiting its reliability, and as we show, commonly used PPs
give disparate results.  In this Letter, we push the state of the art
in accuracy by introducing a method for an all-electron DMC simulation
of solids, eliminating the bias from pseudopotentials, and apply it to
create a high-accuracy pressure calibration scale.
 
The combination of ultra-high pressure mineralogy with seismology has
yielded a wealth of insight into the internal structure of our planet.
Pressure is the key that links these disciplines, mapping phase
transitions and mineral properties to planetary depth.  Establishing
an absolute pressure calibration at multi-megabar pressures,
however, poses a fundamental and continuing problem for high-pressure
experiment.  Current primary calibrations are based on data from
shock-wave experiments, which infer pressure from conservation of
momentum and energy as the shock traverses the sample.  Experimental
uncertainty, combined with errors in model extrapolation, yield scales
which differ from each other by as much to 7\% at room temperature,
with even greater discrepancies at high temperature\cite{Fei07}.  Such
disparity remains a serious obstacle to a quantitative understanding
of Earth's interior.

A pressure calibrant is a material with a known equation of state
(EOS), a small sample of which may be included in hydrostatic
equilibrium with the test subject (e.g. ruby\cite{MaoRuby}).  As an
alternative to shock experiments, high-pressure Brillouin scattering,
which provides a measurement of the bulk modulus, in conjunction with
X-ray diffraction measurements of volume, can be integrated to provide
an EOS, but a correction must be made to transfer from an adiabatic to
an isothermal path\cite{MgOscale}. New approaches such as
quasi-adiabatic Z-pinch based experiments\cite{Zpinch,Remo20081516}
also hold future promise for a primary scale.  There have been
attempts to refine the ruby scale\cite{Holzapfel2005}, and new
calibrations have also been suggested\cite{MgOscale}. Cubic boron
nitride(c-BN) has been identified as a promising material for a new
scale\cite{GoncharovEOS}.  In this Letter, we provide a new pressure
scale based on DMC calculations of the EOS and Raman frequency of
c-BN.  This theoretical approach has the advantage that, in contrast
to present experimental approaches, the method works equally well
under high compression, and uncertainty does not grow with pressure.

Pressure is the negative volume derivative of the Helmholtz free
energy.  In a wide-gap insulator such as c-BN, the free energy can be
written as a sum of the frozen lattice enthalpy, dependent only on
volume, and a phonon thermal free energy, which depends on both volume
and temperature.  Since the static enthalpy is the dominant
contribution at ordinary temperatures, errors in a theoretical EOS can
most often be attributed to the static part.  Previous calculations of
the EOS of c-BN have been based on DFT\cite{GoncharovEOS}, which use
approximate functionals to treat electron exchange and correlation.
Several functionals are in common use, each giving rise to a different
EOS, and there is no {\em a priori} way to predict which will give the
most reliable result.

Quantum Monte Carlo simulation methods explicitly treat electron
exchange and correlation instead of resorting to approximate
functionals.  Variational Monte Carlo (VMC) computes properties by a
Metropolis sampling of a trial wave function.  Diffusion Monte Carlo
samples the many-body ground state of the Hamiltonian through a
stochastic projection of the trial function.  In practice, a fixed
node approximation is used for fermions, such as electrons, which
tests have shown give relatively small error when the nodes are
obtained from high-quality DFT orbitals for electronically simple
materials such as c-BN.  DMC for solids has been demonstrated to give
significantly more accurate cohesive energies\cite{QMC4solids},
equations of state and Raman frequencies\cite{Maezono}, and phase
transitions\cite{Kolorenc} than DFT.  We have used both the CASINO QMC
software suite\cite{CASINO} and QMCPACK\cite{QMCPACK}.

QMC simulations of solids are currently performed within the
pseudopotential (PP) approximation, in which the core electrons are
eliminated and their effect replaced by a nonlocal potential
operator\cite{QMC4solids}.  Since PPs are presently constructed
with a lower-accuracy theory, such as Hartree-Fock (HF) or DFT, this
replacement represents an uncontrolled approximation.  To eliminate
this error, we develop a method for all-electron (AE) QMC simulations
of solids in QMCPACK using trial wave functions derived from
full-potential linearized augmented plane wave (FP-LAPW) calculations
using the EXCITING code\cite{EXCITING}.  Space is divided into
spherical {\em muffin tin} regions around the nuclei, and an
interstitial region.  Orbitals are represented inside the muffin tins
as a product of radial functions and spherical harmonics, and outside
as plane-waves.  To ensure that a wavefunction satisfies the
variational principle, it must be both continuous and smooth.  We
utilize a {\em super}-LAPW formalism that enforces continuity and
smoothness at the muffin tin boundary.  For efficiency, we represent
the orbitals as 3D B-splines in the interstices and the product of
radial splines and spherical harmonics in the muffin tins.

Since AE QMC simulations are computationally expensive, we perform
these simulations in 8-atom cubic supercells.  Simulation cells this
small would typically have significant finite-size errors.  We
eliminate these errors by combining data from AE simulations with that
from PP simulations performed in both 8-atom and 64-atom supercells.
Thus, we are able to simultaneously eliminate systematic errors from
pseudpotentials and from finite-size effects.  The corrected static-lattice 
energy is given at each volume as
\begin{equation}
E = E^{\text{PP}}_{64} + \left[ E^\text{AE}_8 - E^\text{PP}_8 \right]
+ \Delta^\text{MPC}_{64} + \Delta^\text{kinetic}_{64},
\end{equation}
where the term in brackets removes the psuedopotential bias.
$\Delta^\text{MPC}_{64}$ and $\Delta^\text{kinetic}_{64}$ are,
respectively, potential\cite{MPC} and kinetic\cite{Chiesa} corrections
for finite-size errors.  We perform this procedure with three
different PP sets commonly used in QMC: HF PPs from Trail and Needs
(TN)\cite{TrailNeeds1,TrailNeeds2}; HF PPs from Burkatzki, Filippi,
and Dolg (BFD){\cite{BFD}; and DFT-GGA\cite{WuCohen} PPs generated
with OPIUM (WC)\cite{OPIUM}.  Performing the same procedure with
128-atom PP simulations and 16-atom AE simulations yields
statistically indistinguishable results, demonstrating that finite-size
errors are converged at this size.  Additional methodological
details and the finite-size data can be found in EPAPS document
No. \#.

To compute the phonon free energy, we use density functional
perturbation theory (DFPT) in the QUANTUM ESPRESSO package\cite{PWscf}
with the Wu-Cohen functional and the OPIUM PPs.  The phonon density of
states, from which we derive thermodynamic data, is usually very
well-described with DFT.

We compute the free energy for our c-BN system at twelve unit-cell
volumes, spanning volume compression ratios from 0.84 to 2.0,
corresponding to pressures of about -50 GPa to 900 GPa.  While it is
not fully certain that the cubic phase of BN is stable to 900 GPa,
no other structure has been observed, and theoretical studies have
not identified any transitions below 1 TPa\cite{FHK}.  We use the Vinet
form\cite{Vinet} for the isothermal EOS, which we find represents our
free-energy data very well, yielding the bulk modulus, $B_0$, its
pressure derivative, $B_0'$, and the equilibirum volume, $V_0$
(Table~\ref{tab:Parameters}\ref{tab:300KEOS}).  The statistical error
bars for each data point are directly determined from our QMC data.
We compute statistical confidence ranges, taking into account
parameter cross-correlations with a simple Monte Carlo procedure.

\begin{table}
\centering
\subtable[\ Parameters for 300K EOS\label{tab:300KEOS}]{
{\footnotesize
\begin{ruledtabular}
\begin{tabular}{c c c c c }
Source        & PP/AE atoms       & $V_0$ ($\text{\AA}^3$) & $B_0$ (GPa) & $B'_0$ (nondim.) \\\hline
Trail-Needs PP + AE       & 64/8  & 11.792(18) & 381(6) & 3.87(6) \\
OPIUM GGA PP + AE         & 64/8  & 11.769(17) & 385(6) & 3.86(6) \\
BFD HF PP + AE            & 64/8  & 11.781(20) & 382(7) & 3.87(7) \\\hline
BFD HF PP + AE   & 128/16 & 11.812(8)  & 378(3) & 3.87(3) \\\hline
Datchi {\em et al.}\cite{DatchiEOS}  &  & 11.8124    & 395(2) & 3.62(5) \\
Goncharov {\em et al.} & & 11.817(32) & 387(4) & 3.06(15) \\
\end{tabular}\end{ruledtabular}}}\\
\subtable[\ Parameters for thermal pressure\label{tab:ThermalEOS}]{
\centering
{\footnotesize
\begin{ruledtabular}
\begin{tabular}{cccc}
      & $\theta^0_n (\text{K}\, \text{\AA}^{\!-3n})$          & $\alpha_n (\text{K}^{-1}\, \text{\AA}^{\!-3n})$               & $\beta_n(\text{K}\, \text{\AA}^{\!-3n})$ \\ \hline
$n=0$ & 4.836656$\times 10^3$    & 2.598608$\times 10^{-3}$      & -4.869563 $\times 10^3$ \\
$n=1$ & \!\!\!\! -6.929704$\times 10^1$   & -1.200504$\times 10^{-4}$ & 1.763443 $\times 10^2$ \\ 
$n=2$ & \ \  4.634278$\times 10^{-1}$ &  2.789128$\times 10^{-6}$ & -2.186053 $\times 10^0$ \\
$n=3$ & \ \, -1.273468$\times 10^{-3}$&     -2.008994$\times 10^{-8}$ & \ \ 9.303181 $\times 10^{-3}$ \\ 
\end{tabular}
\end{ruledtabular}}}\\
\subtable[\ Parameters for Raman calibration\label{tab:Calibration}]{
{\footnotesize
\begin{ruledtabular}
\begin{tabular}{ccccccc}
$c_0 (\text{cm}^{-1})$ & $c_1(\text{cm}^{-1})$ & $c_2 (\text{K})$ & $b$ & $R_0 (\text{GPa})$  & $R_1(\text{GPa})$ & $R_2(\text{GPa})$ \\\hline
1055.9  & -144.24 & 1497.8  & 3.0155 & 349.87 & 1849.4 & 112.33 
\end{tabular}\end{ruledtabular}}}
\caption{Parameters for the c-BN EOS and Raman calibration.\label{tab:Parameters}}
\end{table}

Figure~\ref{fig:300KEOS} shows the EOS of c-BN at 300 K,
with experimental data from Datchi {\em et al.}\cite{DatchiEOS} and
Goncharov {\em et al.}\cite{GoncharovEOS}, as well as the present work
from simulations with three different PPs.  The residuals in (c) are
derived from DMC simulation with PPs alone, while those in (b) combine
all-electron and PP data.  There is significant discrepancy between
the theoretical curves in (c), suggesting that PP simulation alone
does not provide sufficient accuracy.  Once the PP data is combined
with AE data, as in (b), all the theoretical curves come into good
agreement.  Our theoretical EOS agrees reasonably with that in
Refs. \cite{DatchiEOS} and \cite{GoncharovEOS} within the
experimentally measured pressure range, but the experimental
extrapolation shows significant deviation at high pressure.

\begin{figure*}
\subfigure[\ Corrected EOS]{
\includegraphics[width=0.32\textwidth]{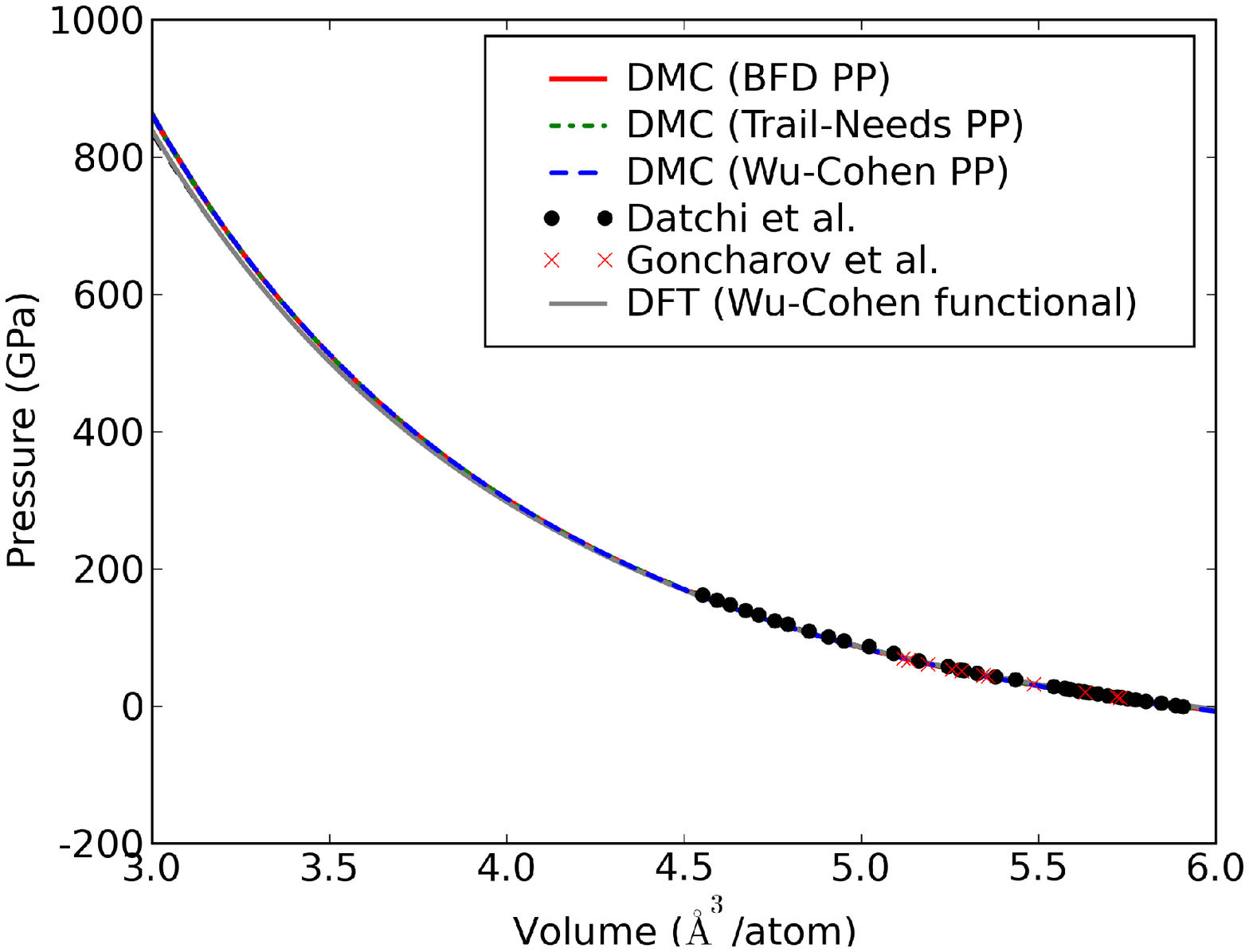}}
\subfigure[\ Corrected pressure residuals]{
\includegraphics[width=0.32\textwidth]{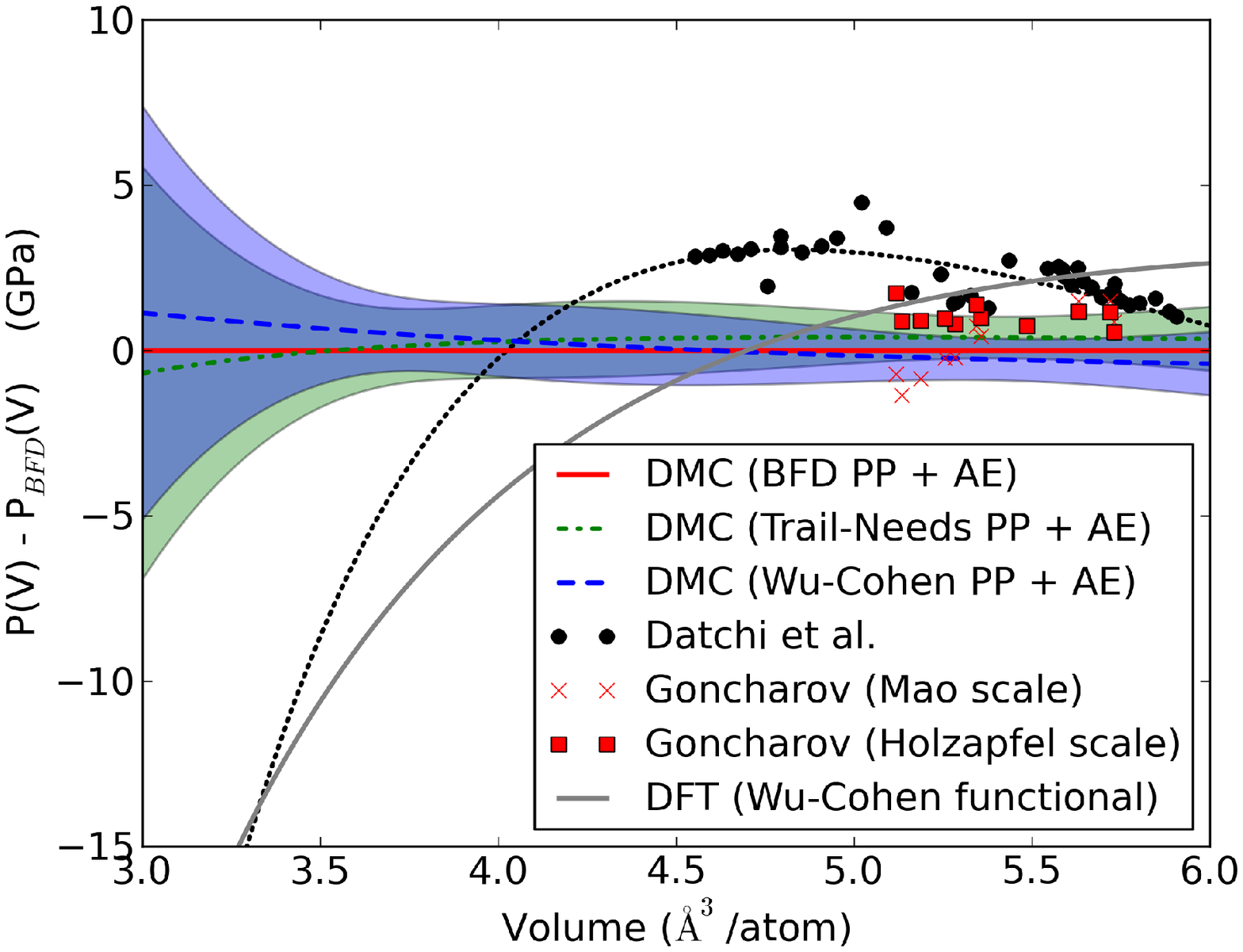}}
\subfigure[\ Uncorrected pressure residuals]{
\includegraphics[width=0.32\textwidth]{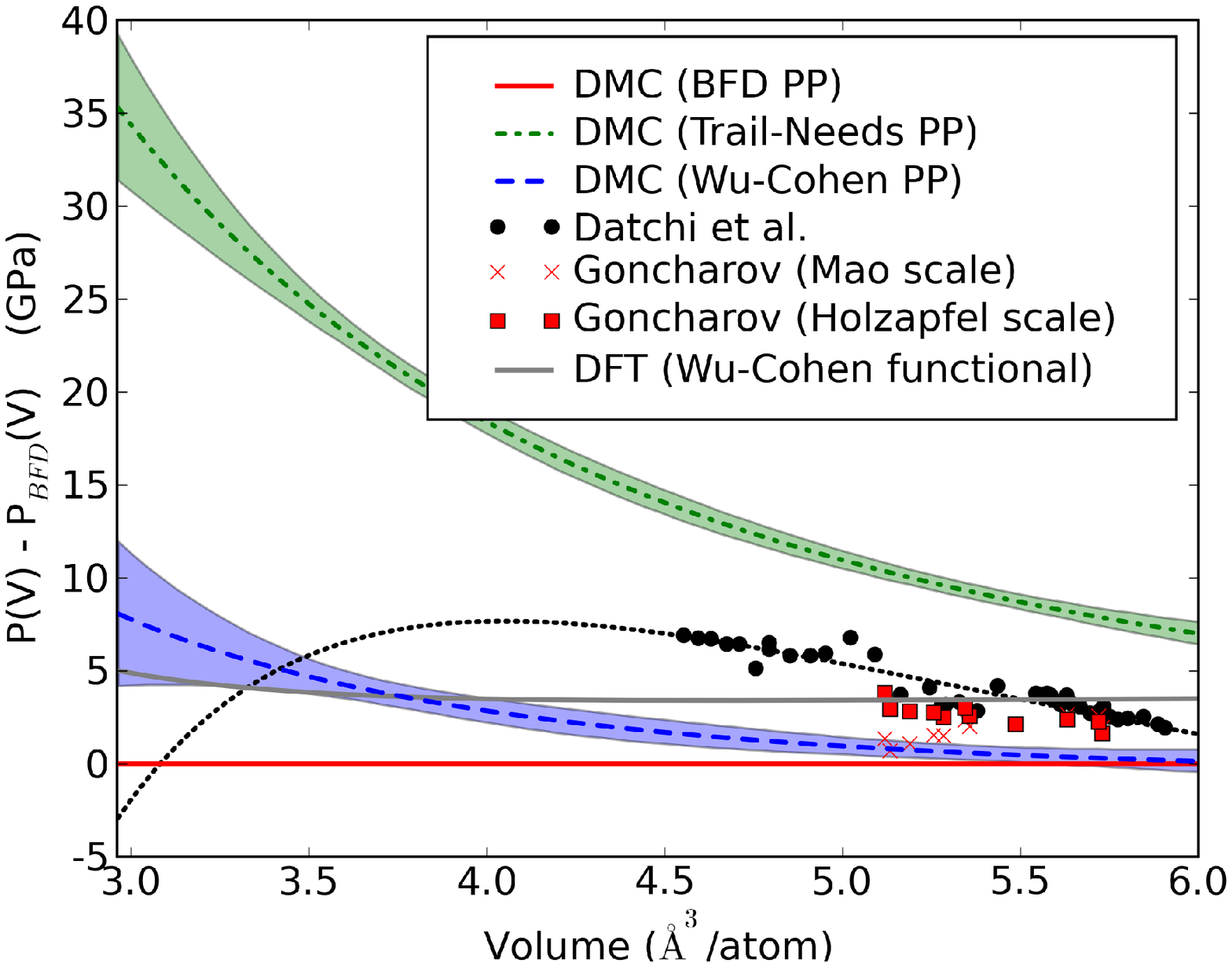}}
\caption{The c-BN EOS at 300 K, computed with DMC, compared to
  experiment.  We plot the full EOS and the pressure residuals with
  respect to DMC with the BFD PP.  (a) gives the corrected EOS
  resulting from combining the EOS and PP data, while (b) gives the
  pressure residuals for the same data.  (c) gives the uncorrected
  pressure residuals from the PP simulations only.  Shaded areas represent
  the one-$\sigma$ statistical confidence region from QMC
  data.\label{fig:300KEOS}}
\end{figure*}

We may write the thermal equation of state in the form
\begin{equation}
P(V,T) = P_{300K}(V) + P_\text{th}(V,T) -  P_\text{th}(V,T=300),
\end{equation}
where $P_{300K}$ is the room-temperature contribution, fit to the Vinet
form.  The phonon contribution is written in an augmented
Debye model,
\begin{eqnarray}
P_\text{th}(V,T) & = & -\frac{\partial F_D(\theta,T)}{\partial \theta} \frac{\partial \theta}{\partial V}\\
\theta(V,T) & = & \theta^0(V) + \beta(V) \exp(-\alpha(V) T) \\
x(V) & = & x_0\! +\! x_1 V\! +\! x_2 V^2\! +\! x_3 V^3, \ x \!\in\!\{\theta^0\!,\alpha,\beta\}
\end{eqnarray}
in which the Debye temperature, $\theta$, is a function of both $V$
and $T$ (Table~\ref{tab:Parameters}\ref{tab:ThermalEOS}).  The Debye free energy per two-atom cell, excluding the zero-point term, is given by
\begin{equation}
F_D = 6 k_B T
\left[\ln\left(1-e^{\frac{\theta}{T}}\right)-\left(\frac{\theta}{T}\right)^{\!\!3}\int_0^{\frac{\theta}{T}}
\frac{x^3}{e^x-1} dx\right].
\end{equation}

c-BN can be used to calibrate pressure optically by measuring the
frequency shift of the TO Raman mode, allowing bench-top experiments.
We compute the pressure and temperature dependence of this frequency.
Within the quasiharmonic approximation, phonon frequencies have
explicit dependence on volume only.  At constant pressure, however,
these frequencies have an implicit temperature dependence resulting
from thermal expansion.  The dependence due to thermal expansion
accounts for only about half the total $T$-dependence of the Raman
mode, as observed in \cite{DatchiRaman} and \cite{GoncharovRaman}.
The remaining $T$-dependence can be attributed to significant
anharmonic effects in c-BN, and is included in our calculations.

Since the optical branch has small dispersion, we treat the
anharmonicity as a one-dimensional on-site anharmonic oscillator, in a
similar approach to the QMC computation of the TO Raman frequency of
diamond\cite{Maezono}.  At each volume, we compute the effective
Born-Oppenheimer potential well for the TO mode with DMC and the BFD
PP at nine displacements along the mode eigenvector in the 64-atom
supercell.  We fit the data to a quartic polynomial, and numerically
solve the 1D Schr\"{o}dinger equation in this analytic potential.
This results in a set of single-phonon energy levels, $\{E_n\}$, with
nonuniform separation.  From $\{E_n\}$, we compute an
intensity-weighted average Raman frequency, $\bar{\nu}$, as a
function of pressure and temperature.  The matrix element for the
transition from $n$ to $n-1$ is proportional to $\sqrt{n}$ and is
thermally weighted by the Boltzmann occupation of state $n$, so
the intensity-averaged frequency, $\langle\nu\rangle$, can be
given as
\begin{equation}
\langle\nu\rangle  =  \frac{\sum_{n=1}^\infty I_n\frac{E_n - E_{n-1}}{hc}}{\sum_{n=1}^\infty I_n}, \ \ \ \text{where} \ \ \ 
I_n  =  n\, e^{-\frac{E_n}{k_B T}}.
\end{equation}
The excess thermal softening, i.e. beyond that from thermal expansion
alone, is accounted for by the thermal average of the anharmonic
frequencies.  Figure~\ref{fig:RamanCalibration} shows the computed
Raman frequencies compared with the experimental data reported in Refs. \cite{DatchiRaman,GoncharovRaman,GoncharovHPR}.

\begin{figure}
\subfigure[\ Fit to present theory for Raman shift.\label{fig:RamanFit}]{
\includegraphics[width=0.50\textwidth]{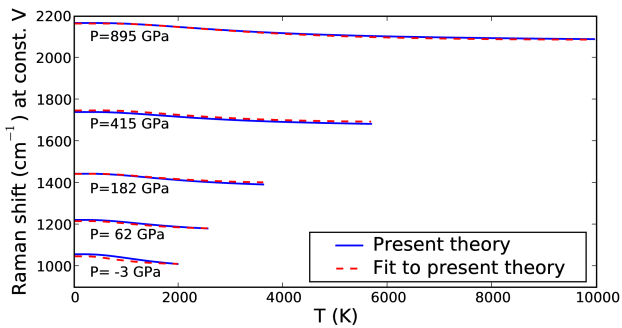}}
\subfigure[\ Comparison of Raman experiment and theory\label{fig:RamanComp}.]{
\includegraphics[width=0.448\textwidth]{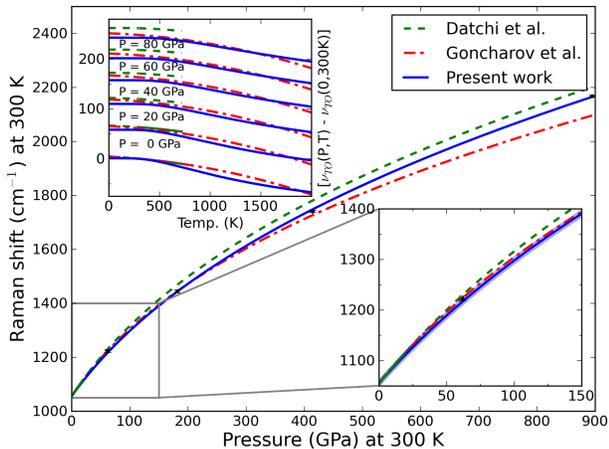}}
\caption{TO Raman frequency of c-BN as a function of temperature and
  pressure.  (a) gives compares our theoretical Raman frequency with
  the fitted form in Eqs.~\ref{eq:RamanFit1} and \ref{eq:RamanFit2}.  (b)
  compares our calibration at 300K with the experimental results from
  \cite{DatchiRaman} and \cite{GoncharovRaman}.  The black dots with
  error bars are the QMC frequencies, while the blue solid line gives
  the fit.  The lower inset gives an expanded view at low pressure, in
  which the shaded region gives the statistical confidence region.
  The upper inset gives the variation with temperature.\label{fig:RamanCalibration}}
\end{figure}

Both Refs. \cite{DatchiRaman} and \cite{GoncharovRaman} give a
ruby-like calibration formula, which can be expressed as
\begin{equation}
P = (R/b)\left[ (\nu/\bar{\nu})^b - 1\right],
\end{equation}
where $R$, $\bar{\nu}$, and $b$ have quadratic $T$-dependence.  This
dependence is sufficient below 2000 K, but cannot represent our data
at high temperature.  We use a form which captures the Boltzmann
occupation of phonon excitations,
\begin{eqnarray}
\nu(P,T) & = & \nu_0(P) + \nu_1(P) \exp\left[-\frac{\nu_2(P)}{T}\right] \label{eq:RamanFit1}\\
\nu_n(P) & = & c_n \left[\frac{bP}{R_n} + 1\right]^{1/b}\!\!\!\!\!\!, \ \ \ \ n=1,2,3 \label{eq:RamanFit2}
\end{eqnarray}
with parameters in Table~\ref{tab:Parameters}\ref{tab:Calibration} and
plotted in Figure~\ref{fig:RamanFit}.  Note that this formula cannot
be analytically inverted, but a very simple iterative solution can be
used for calibrating pressure from $\nu$ and $T$.

The main axis of Figure~\ref{fig:RamanComp} gives the room-temperature
Raman frequency versus pressure.  There is good agreement in the
relatively low-pressure region in which the Raman frequency was
measured.  At very high pressure, the deviation with respect to the
extrapolation in \cite{DatchiRaman} increases with a maximum
discrepancy of 38 cm$^{-1}$ or, conversely, a deviation in the
pressure calibration of 50 GPa at 900 GPa.  The deviations with
respect to \cite{GoncharovRaman} are 70 cm$^{-1}$ and 120 GPa.  The
experimental parameters capture the correct qualitative high-pressure
behavior up to 900 GPa, despite the fact that data was available only
to 20 and 64 GPa, respectively.  This suggests the form for the fit
was well-chosen.

We have presented a fully {\em ab initio} pressure calibration based
on quantum Monte Carlo simulations, and have introduced a method for
all-electron simulations of solids to eliminate bias from
pseudopotentials.  This method should be applicable to at least
first-row solids, allowing increased accuracy in the study of other
materials and providing a new benchmark for other methods.  The only
remaining systematic error in the static contribution to the EOS is
from the fixed-node approximation used in DMC.  For simple materials
such as c-BN this error should be quite small, and tends to cancel
between different volumes.  Thus we believe the EOS is robust enough
to be used directly in experiment as a primary pressure calibrant, and
can be used to cross-calibrate scales based on other materials.  Since
the accuracy of our methods should not depend on compression,
our calibration can be used up to 900 GPa.

This work was supported under the National Science Foundation Grant
EAR-0530282.  This research used resources of the National Center for
Computational Sciences and the Center for Nanophase Materials
Sciences, which are sponsored by
the U.S. Department of Energy.  
This work was partially supported by the National Center for Supercomputing
Applications under MCA07S016 and utilized the Abe and Lincoln clusters.
We thank Alexander Goncharov, F. Mauri and M. Lazzeri for helpful 
suggestions and discussions.

\end{document}